\documentstyle[pra,aps,twocolumn,floats]{revtex}
\begin{document}


\draft

\wideabs{

\title{The validity of the Landau-Zener model for output coupling
of Bose condensates}

\author{J.-P. Martikainen and K.-A. Suominen}

\address{Helsinki Institute of Physics, PL 9, FIN-00014 Helsingin yliopisto,
Finland}

\date{\today}

\maketitle

\begin{abstract}
We investigate the validity of the Landau-Zener model in describing the
output coupling of Bose condensates from magnetic traps by a
chirped radiofrequency field. The predictions of the model are compared
with the numerical solutions of the Gross-Pitaevskii equation. We find a
dependence on the chirp direction, and also quantify the role of gravitation.

\end{abstract}

}

\narrowtext

Bose-Einstein condensation of alkali atoms in magnetic traps was first observed
in 1995~\cite{Anderson95}. The first step towards coherent matter beams
was taken when a controlled release of condensed atoms from a magnetic trap was
demonstrated in MIT~\cite{Mewes97}, and recently at Garching~\cite{Bloch99}. 
The basic tool for output coupling has
been spin-flipping induced by a radiofrequency (rf) magnetic
field~\cite{Mewes97,Bloch99}, but other approaches have also been
used~\cite{Anderson98,Hagley99}. 

The MIT group demonstrated output couplings based on chirping the rf field, and
on resonant rf pulses~\cite{Mewes97}. In both cases the experimental results
are understood in terms of simple models. The theory of the pulsed and cw 
output couplings has been studied extensively in the
literature~\cite{Ballagh97,Naraschewski97,Jackson98,Graham99,Band99}.  
In this
Brief Report we examine the chirped output coupler. We find the validity
conditions for the multistate Landau-Zener (MLZ) description~\cite{Vitanov97}
and compare them with the numerical solutions of the Gross-Pitaevskii (GP)
equation~\cite{Nozieres90}. We chart the combinations of the chirp speed
$\lambda= d\omega_{\rm rf}/dt$ and the rf field amplitude $\Omega$ for which the
MLZ description fails. We show that this failure depends on the chirp 
direction. Also, the repulsion between particles in the condensate and
gravitation are found to contribute to the validity conditions. 

\begin{figure}
\vspace*{40mm}
\includegraphics{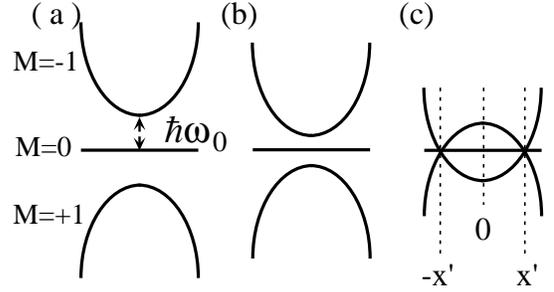}
\caption[f1]{Position dependence of the energy levels of atoms in a
magnetic trap. (a) No output coupler, (b) output coupler with
$\omega_0>\omega_{\rm rf}$, (c) output coupler with $\omega_0< \omega_{\rm
rf}$.\label{fig1}}
\end{figure}

Magnetic trapping is achieved by spin-polarising the atoms. For simplicity we
consider the case of $F=1$, where $F$ is the hyperfine quantum number. In
the inhomogeneous magnetic field the three substates $M=-1,0,1$
experience the potentials shown in Fig.~\ref{fig1}(a). The atoms on the $M=-1$
state are trapped in the harmonic potential. In the MIT trap one
had $\omega_x=\omega_z=(2\pi)\ 320$ Hz and $\omega_y=(2\pi)\ 18$
Hz~\cite{Mewes96}. When the spin of an atom is flipped from this state to the
$M=0$ state, the atom moves away due to quantum mechanical dispersion and the
repulsion between the condensed atoms. If the internal state of the atom is
changed into the $M=1$ state, it feels the inverted parabolic
potential as well. 

We can eliminate the rf oscillations by making the rotating wave
approximation, and then shifting each $M$ state in energy by an appropriate
number of photons~\cite{Ketterle96}. Now the field-induced resonances appear as
potential crossings. 
At the rf field frequency
$\omega_0$ the atoms at the center of the trap are in resonance with the field.
We define the field detuning as $\Delta=\omega_{\rm rf}-\omega_0$. For
$\Delta<0$ none of the atoms are in resonance, and for $\Delta>0$ atoms at
locations $x=\pm x'$ are in resonance [Fig.~\ref{fig1}(b) and
(c)]. 

In the chirped output coupler one sweeps $\omega_{\rm rf}$ so that all atoms in
the trap feel a resonant field for a brief moment. The idea is to make
this moment long enough for achieving a total or partial spin flip, and brief
enough that the atoms do not have time to move due to changes in their internal
state. Then the atoms remain stationary while they experience a time-dependent
change of the energy difference between adjacent spin states. 

In the MIT experiment a linear chirp was used: $\Delta=\lambda t$. For two
internal states this corresponds directly to the Landau-Zener (LZ) model. In
moderate magnetic fields the Zeeman shifts are linear and thus all adjacent
states are resonant simultaneously, so one has a genuine multistate
problem~\cite{Vitanov97,Ketterle96}. In this particular case, however, the spin
dynamics of a stationary atom can be described analytically, by a simple
multistate generalization of the Landau-Zener result (MLZ)~\cite{Vitanov97}. In
the experiment the agreement with the MLZ prediction was good. 

For simplicity we consider one spatial direction only. The three-state
Hamiltonian for a stationary atom located at $x$ is
\begin{equation}
   H(x,t) = \left(\begin{array}{ccc} \frac{1}{2}m\omega^2x^2-\hbar\Delta &
            \frac{\hbar\Omega}{\sqrt{2}}& 0 \\
            \frac{\hbar\Omega}{\sqrt{2}} & 0 & 
            \frac{\hbar\Omega}{\sqrt{2}} \\
            0 & \frac{\hbar\Omega}{\sqrt{2}} &
            \hbar\Delta-\frac{1}{2}m\omega^2x^2
            \end{array}\right).
\end{equation}
Here $m$ is the atomic mass, $\omega$ is the trap frequency and $\hbar\Omega$
is the rf field coupling. For a stationary atom we can solve the
time-dependent Schr\"odinger equation with $H$. The MLZ theory predicts the
final populations of the spin states:
\begin{eqnarray}
    P_{-1}&=&\exp(-2\pi\Gamma),\\
    P_0&=&2\exp(-2\pi\Gamma)[1-\exp(-2\pi\Gamma)],\\
    P_{+1}&=&[1-\exp(-2\pi\Gamma)]^2, \label{PLZ}
\end{eqnarray}
where $\Gamma=\Omega^2/(2 \lambda)$~\cite{Mewes97,Vitanov97}. In the MIT
experiment $\lambda=(2\pi)\ 500$ MHz/s, and $\Omega$ changed from 0 to about
$(2\pi)\ 11.3$ kHz. The $P_i$'s depend only on $\Gamma$ and not on the trap
geometry; thus in asymmetric traps the output coupling takes place in all
directions with the same efficiency.

In the Gross-Pitaevskii theory the single atom amplitudes $\Psi_i(x,t)$ describe
effectively the whole condensate (i.e., $|\Psi_i(x,t)|^2$ gives the density
distribution of the atoms on spin state $i$)~\cite{Nozieres90}. Their time
evolution is obtained from the time-dependent Schr\"odinger equation with the
Hamiltonian
\begin{equation}
	{\cal H}_{i,j}=	-\frac{\hbar^2}{2m}\frac{\partial^2}
                 {\partial x^2}\delta_{i,j}+H_{i,j}+
                 \sum_{l=1}^3C_{i,l}|\Psi_l|^2.
\end{equation}
The last term describes interactions between the particles. The parameters
$C_{i,l}$ are proportional to atom numbers and the scattering lengths of the
corresponding internal states $i$ and $l$. In our one-dimensional 
study this parameter does not match properly with the realistic three 
dimensional situation. 
But a trap with a very low frequency at one direction can be regarded quasi
one-dimensional. In this case a reasonable estimate for our 1D $C$ parameter
would be $C=C_{3D}/A$, where $A$ is the cross sectional condensate area. As an
order of magnitude estimate it should be valid for other traps as well.
For simplicity we take all $C$'s to be equal.

We solve the GP equation numerically for the output coupler. 
In the limit of large
$C$ we can ignore the kinetic energy term and obtain the Thomas-Fermi
solution, $\Psi(x,0)=\sqrt{[\mu-U(x)]/C}$, where $U(x)$ is the trapping
potential. The condition $\mu-U(x)\geq 0$ defines the edge of the condensate.
The chemical potential $\mu$ is obtained from the normalisation of the wave
function. 

A breakdown of the MLZ model is expected if the atoms move during the
transition process. A similar problem arises for  diatomic molecules
interacting with short laser pulses~\cite{Garraway95}. In order to quantify
this breakdown we consider the characteristic time scale
$\delta t$ of the LZ process,  $\delta t\simeq\Omega/\lambda$~\cite{Vitanov99}.
The atoms need to remain stationary during this time. The term "stationary" can
be defined by transforming $\delta t$ into a region $\delta x$ around the
location $x_0$ of the atom. For simplicity we assume that $x_0>0$. For
parabolic potentials in the $F=1$ case we set $\hbar\Omega= \frac{1}{2}m
\omega^2 [(x_0+\delta x)^2-x_0^2] \simeq m\omega^2x_0 \delta x$, which defines
$\delta x$. If the atom moves a distance $\Delta x$ in time
$\delta t$, it can be regarded stationary if $\Delta x\ll \delta x$.

The atomic motion can arise from quantum mechanical diffusion, repulsion
between atoms, or from acceleration along the inverted parabolic potential. We
consider the acceleration $a$ first. With Newtonian dynamics we get $a(x_0)=
-(1/m) (\partial U/\partial x) |_{x=x_0}= \omega^2x_0$. In the small region
around $x_0$ we have $\Delta x = a(\delta t)^2/2 = \omega^2x_0 \Omega^2/
(2\lambda^2)$. Thus we get the condition $\lambda^2 / \Omega\gg
m\omega^4x_0^2/\hbar$. This needs to be true for all $x_0$; the right-hand side
is maximised at the edge of the condensate. 

For small $C$ the edge is near the width of the ground state of the
harmonic potential, max$(x_0) \simeq \sqrt{\hbar/m\omega}$, which gives
\begin{equation}
   \frac{\lambda^2}{\Omega\omega^3}\gg 1.\label{condition1}
\end{equation}
For large $C$ we take the Thomas-Fermi approximation, and then max$(x_0) \simeq
\sqrt{2\mu/m\omega^2}$, which gives
\begin{equation}
   \frac{\lambda^2}{\Omega\omega^2}\gg \omega_{\mu},\label{condition2}
\end{equation}
where $\mu=\hbar \omega_{\mu}$. The MIT trap parameters satisfy condition
(\ref{condition1}) well in all directions, and to break down condition
(\ref{condition2}) would require an unrealistically large $\mu$. Since
$\lim_{C\rightarrow 0}\mu=\hbar\omega/2$, Eq.~(\ref{condition1})
is a special case of Eq.~(\ref{condition2}).

Diffusion and repulsion can give atoms a velocity which initially overcomes 
the acceleration. For the $M=0$ state these processes are naturally covered by
the various studies of the ballistic expansion of condensates; see
Ref.~\cite{Parkins98} and references therein. As we are only
looking for constraints it is sufficient to characterise the maximum
speed $v$ of the atoms with the energy stored in the trapped condensate. We set
$\mu=mv^2/2$. For small $C$ the speed $v$ reduces to the free-space
momentum width of the Gaussian harmonic oscillator wave function, 
$v\simeq \sqrt{\hbar\omega/m}$. Now $\Delta x\simeq v \delta t$. On the other 
hand, at $x_0=0$ we have $\delta x= \sqrt{2\hbar \Omega/ (m\omega^2)}$.
Eventually we get for diffusion/repulsion the conditions (\ref{condition1}) 
and (\ref{condition2}).

These conditions do not depend on the direction of the chirp. However,
there exists another breakdown mechanism for the MLZ theory. Atoms that have
interacted resonantly with the field can re-enter the resonance region (a
reunion) due to their motion. Let us consider a positive chirp: a
resonance emerges at $x=0$ and then separates into two points that move towards
large $|x|$. This is demonstrated in Fig.~\ref{fig1} if we consider it as 
a sequency of snapshots. Strong acceleration on the $M=1$ state leads to the 
above problem. We need a condition on $\lambda$ and $\Omega$ for avoiding 
the reunion until the transition probability is negligible.

Here the direction of the chirp is crucial. For negative chirps the resonances
emerge at large $|x|$ and move towards $x=0$ where they disappear.
Acceleration, however, moves the atoms in the opposite direction. Thus for
negative chirps the reunion problem is absent. The role of chirp direction has
been studied e.g. in the context of laser-molecule
interactions~\cite{Garraway95}.

We consider the case where the reunion happens at $x=x_0$. Energy
conservation gives us roughly the local speed $v_0$ of the moving atoms:
$v_0\simeq \omega x_0$. The slope of the local energy difference between
adjacent states is $\alpha_0=m\omega^2x_0$. The product of these quantities
gives us the local motion-induced change in the energy levels for the atom:
$\hbar\lambda_0 = |dU/dt| =|(\partial U /\partial x) (\partial x/ \partial t)
|_{x=x_0} =m\omega^3x_0^2$. The motion of the resonance, $\lambda$, is small
compared to $\lambda_0$ for realistic parameters, and we can ignore it.
Basically, we want that the motion-induced LZ transition probability
at $x_0$ is smaller than some fixed value $\tilde{\Gamma}$:
\begin{equation}
   \Gamma=\frac{\Omega^2}{\lambda_0}\leq\tilde{\Gamma}.\label{condition3}
\end{equation}

The time it takes for a resonance to reach $x_0$ is $t_0=m\omega^2 x_0^2/(2
\hbar\lambda)$. For a accelerating atoms we need to solve the Newtonian
equation of motion: $(\partial^2 x /\partial t^2) -\omega^2x =0$. For
$x'=\sqrt{2\hbar\lambda t'/(m\omega^2)}$, $v(t')=v'$ we get 
\begin{equation}
   x=x'\cosh[\omega (t-t')]+\frac{v'}{\omega}\sinh [\omega (t-t')].
   \label{solutionx}
\end{equation}

\begin{figure}[htb]
\vspace*{12cm}
\includegraphics{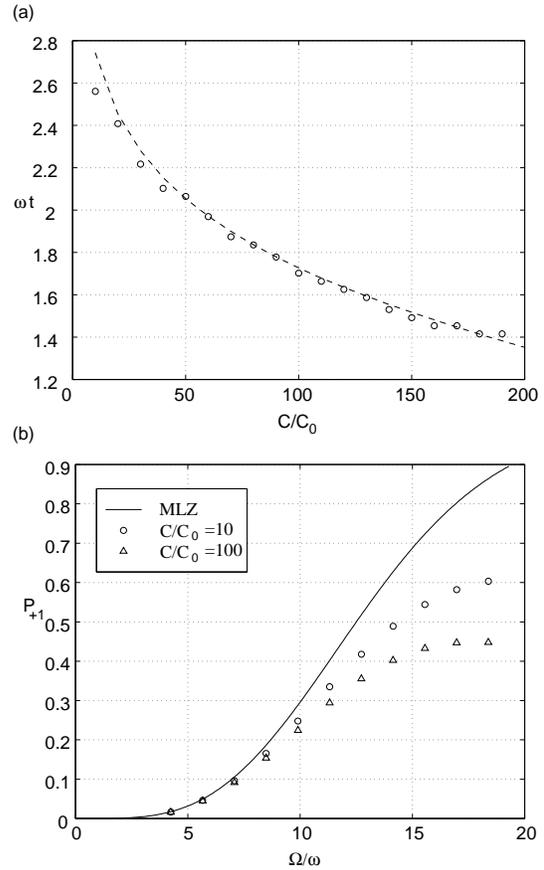}
\caption[f2]{(a) The time it takes for the spin-flipped atoms to reach the
reunion as a function of $C$, and when and $\lambda=100\omega^2$. 
The circles are the GP results, and the dashed
line is the classical estimate based on Eq.~(\ref{solutionx}). (b) The
breakdown of the MLZ theory as a function of $\Omega$ for two values of $C$,
and $\lambda=200\omega^2$. Here $C_0=\sqrt{\hbar^3\omega/m}$. \label{fig2}}
\end{figure}

There are several possible values for $x'$ and $v'$, and the quantum mechanical
diffusion/repulsion complicates this simple Newtonian picture. For large
$C$ we can assume that the particles which reach the reunion first come from
the edge of the initial condensate. This fixes $x'$ (and $t'$). We have
simulated the problem numerically with the GP equation, and obtained the reunion
time for the fastest atoms. As shown in Fig.~\ref{fig2}(a) an effective
trajectory $v'\propto C^{1/3}$ locates the reunion well. The main dependence on
$C$, however, arises from the location of the condensate edge, $x'\propto
C^{1/3}$. Some examples of the breakdown are shown in Fig.~\ref{fig2}(b).

\begin{figure}
\vspace*{62mm}
\includegraphics{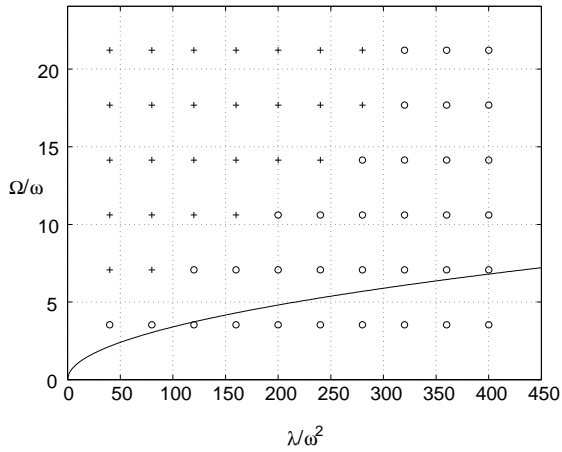}
\caption[f3]{The validity of the MLZ theory. The model predictions
for $P_{+1}$ are compared with the GP solutions for the case of
positive ($\lambda>0$) chirp direction. Here $C=100C_0$, 
which corresponds well to the Thomas-Fermi
limit. The plus signs ($+$) indicate when the values differ by more
than 10 \%, and circles when less. In the region below the solid
line the reunion is not expected to produce more than 10 \%
deviation even if all atoms on the $M=+1$ state would reach the
reunion as soon as the first ones do it in simulations.\label{fig3}}
\end{figure}

One should not make detailed conclusions from such trajectories. Apart from
the crudeness of the Newtonian mechanics, the fastest atoms are only a fraction
of the condensate. In Fig.~\ref{fig3} we show in the ($\lambda,\Omega$) plane
where the MLZ prediction for $P_{+1}$ fails for more than 10 \%\ for the 
large $C$ case. The constraint obtained by using the effective trajectory and 
condition~(\ref{condition3}) is clearly too demanding. Also, in a real 
experiment one can switch the field off before the reunion and thus avoid the 
problem. This is the case in the MIT experiment: after reaching the resonance 
$\omega_0$ the rf field is on for about $\tau\simeq 0.5$ ms. As $\tau 
\omega_x=1$ and $\tau \omega_y =0.06$, the field is off by the time of reunion,
as Fig.~\ref{fig2}(a) shows. For a negative $\lambda$ we saw no large deviation
from the MLZ prediction for the values used in Fig.~\ref{fig3}.

\begin{figure}[hbt]
\vspace*{55mm}
\includegraphics{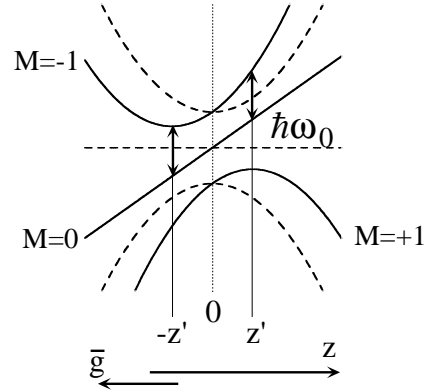}
\caption[f4]{Position dependence of the energy levels of atoms in a
magnetic trap under the effect of gravitation. The shift factor for 
the potential centers is $z'=g/\omega^2$. The dashed lines show
the magnetic trap potentials without the effect of gravitation
$\vec{g}$.\label{fig4}}
\end{figure}

The effects of gravitation should be considered in output
couplers~\cite{Graham99,Jack99}. In the direction of gravitation ($z$) 
the trapping potentials are
\begin{eqnarray}
   V_{-1} &=& \frac{1}{2}m\omega_z^2\left(z+\frac{g}{\omega_z^2}\right)^2
              -\frac{mg^2}{2\omega_z^2},\\
   V_0    &=& mgz\\
   V_{+1} &=& -\frac{1}{2}m\omega_z^2\left(z-\frac{g}{\omega_z^2}\right)^2
              +\frac{mg^2}{2\omega_z^2},
\end{eqnarray}
where $g$ is the gravitational acceleration. The $M=\pm 1$ potentials remain
harmonic but have spatially shifted centers, as shown in Fig.~\ref{fig4}. 
As gravitation 
affects only the external degrees of freedom the rf field resonance conditions 
still follow the purely magnetic potentials. In other words, the resonance 
points are located symmetrically in respect to the magnetic field minimum, 
but not in respect to the trap center. For a stationary atom the LZ parameter 
$\Gamma$ remains unaffected by gravitation, but many other conditions change.

For the MIT trap one gets $(\Delta x)_g=g/\omega_z^2=2.5\ \mu$m which is about
the size of the trap ground state (1.7 $\mu$m) but clearly smaller than the
condensate (17 $\mu$m). For gravitation to dominate acceleration on the $M=1$
state at the condensate edge we obtain the condition $2\mu \omega_z^2<mg^2$ 
(for large $C$), which reduces to $\hbar\omega_z^2<mg^2$ as $C\rightarrow 0$. 
If gravitation dominates, then the basic validity condition for the MLZ 
approach becomes ($\Delta z =g\Omega^2/\lambda^2$, $\delta z =
\hbar\Omega/(m\omega_z^2z_0)$, $z_0\simeq g/\omega_z^2$)
\begin{equation}
   \frac{\lambda^2}{\Omega}\gg\frac{mg^2}{\hbar}.
\end{equation}
This applies also for the motion on the $M=0$ state.

Figure~\ref{fig4} shows also that the shifts in the
potential centers can be used for directed output coupling. With a very
slow negative chirp one can leak the condensate from the earthside edge of the
trap, as the chirped rf pulse becomes resonant there first. Due to the slowness
the condensate edge will follow the resonance, and the other
resonance point will not reach the diminishing condensate. This approach is
used in the recent experiment~\cite{Bloch99}. It is complementary to
the situation considered by us, as there the chirp timescale must
be clearly longer than the motional time scales. 

In this Brief Report we have derived the validity conditions of the multistate
Landau-Zener approach for chirped output couplers, with and without the 
presence of gravitation. Comparison with numerical results shows that these 
conditions allow one to achive "safe" parameter regions easily in the
experiments. 
Our study is for one dimension only, but our results should apply directly
to three dimensions, especially in the case of strongly asymmetric traps,
where the tightest trapping direction will dominate the expansion of the
untrapped atom cloud.

We acknowledge the support by the Academy of Finland.

\end{document}